\documentclass[a4paper,11pt]{article}
\pdfoutput=1 

\usepackage{jheppub} 

\usepackage{slashed}
\usepackage{amsmath}
\usepackage[numbers]{natbib}
\usepackage{filecontents}

\title{Bosonic und Fermionic Quasinormal Modes of Rotating Black Holes Through AdS/CFT}

\author[a]{Jayant Rao,}
\author[b]{Shubhanshu Tiwari}

\affiliation[a]{ETH Zurich, Zurich}
\affiliation[b]{University of Zurich, Zurich}

\emailAdd{raor@ethz.ch}
\emailAdd{stiwari@physik.uzh.ch}

\abstract{We study the holographic duals of bosonic and fermionic fields and compute their Quasinormal Frequencies (QNF) by exploiting that they are massive charged fields in $\mathrm{AdS_2 \times \mathcal{K}}$. In particular, we assess the stability of the scalar and fermionic operators in Myers-Perry's black holes with equal angular momenta. We compute the central charge and the Hawking-entropy and introduce extremality parameters to determine the properties of the dual CFT. Then, we briefly discuss how detection of Spin-2 fields can pave the way for more deep understanding of the holographic principles. We end with a brief discussion of the implications onto Strong Cosmic Censorship.}

\begin{document}
\maketitle
\flushbottom

\section{Introduction}

Black hole solutions have been of most interest in the context of general relativity and specifically also on the road to understanding how a possible theory of quantum gravity may look like. The investigation of Quasinormal Modes (QNM) has been of particular interest for a long time because it reveals insights on black hole formation, ringdown as well as possible hints on quantum effects \cite{cardoso}. Close to the horizon it can be shown that scalar fields in near extremal black holes behave similar to charged scalar fields subject to a $U(1)$ gauge field on $AdS_2$ \cite{reall}. Thus, it interesting to see what happens in the context of the holographic duality principle to a charged scalar field close to the horizon. \newline
One of the most active research topics in theoretical physics has been the \textit{Holographic Duality}  hypothesis \cite{maldacena} which is best understood now in the form of the AdS/CFT correspondence. Since it is of interest to investigate higher-dimensional generalisations of black holes, it has become very natural to consider the generalisations of the \textit{Kerr-Solution} in higher dimensions which is known as the Myers-Perry black hole. In order to exploit some symmetry effects (but still retaining the main physical information) it has also become advantageous to consider the 1-cohomogeneity solution with equal angular momenta. \newline
Using methods from the aforementioned holographic duality, it has become particularly advantageous to consider the linear response formalism within this theory and then determine the Quasinormal Frequencies which in turn of course correspond to excited states of dual bulk operators. By separating the variables of the different equations of motion, we identify the radial asymptotic behaviour as well as the radial angular eigenvalues. Since we are interested in well-defined dual correlators, we can make a well-defined ansatz for the bulk fields and then use the retarded Green's function in order to find the quasinormal modes with the \textit{correct} boundary conditions. We can then understand using the conformal weights and from there the stability conditions how the conformal Virasoro algebra must look like. We investigate scalars, fermions and tensors in order to understand the stability of such fields and if we can use the holographic dictionary to also make statements on the classical predictions of General Relativity. \newline
Finally, we also investigate tensorial modes close to the horizon and want to understand how such (now detectable) higher spin perturbations look close to the horizon and close to extremality. Specifically, it is interesting to see how the concept of overtone modes comes to play with holography and how we can understand more about the dual CFT from that in terms of the ringdown process. \newline

\section{Bosonic Fields}

In this section, we aim to explain how to use the conjectured holographic duality between $AdS_2$ and a boundary conformal field theory (CFT) in order to find the Quasinormal frequencies (QNFs) of bosonic fields in the near-horizon region (NH) for rotating black holes close to extremality. We then apply our theory to explicitly determine the QNFs for specific solutions of interest to us, foremost the 1-cohomogeneity solution to the Myers-Perry's black hole and secondly the Kerr solution. We are specifically interested in understanding how the NH geometry can be understood in terms of a charged scalar field in the near-horizon case. All of this is only valid inside the near-extremal approximation of the black hole. \newline

\subsection{General Case}
\subsubsection{Preparations}

Our studies are fundamentally motivated by the results obtained by H.Reall and H.Kunduri in which they prove that the NH geometry of a black hole at extremality can be understood to be locally isometric to $AdS_2 \times \mathcal{K}$. Here $\mathcal{K}$ refers to a compact topological manifold. In explicit calculations, we one can directly see how this compact space looks in detail, but we will in general use it to introduce a Kaluza-Klein decomposition of the scalar field. \newline
The general isometry to $AdS_2 \times \mathcal{K}$ allows to write the metric in form of a fibration of an $AdS_2$ metric, such that (in the case of a rotating solution), we can write the most general form of the near-horizon metric as \cite{Kunduri2007NearhorizonSO}:

\begin{eqnarray}
\label{general}
            ds^2 = L(y)^2 \left( -R^2 dT^2 + \frac{1}{R^2} dR^2 \right) +  
        &g_{IJ} (y) (d \phi^I - k^I R dT)  
         (d \phi^J - k^J R dT) +  g_{AB}(y) dy^A dy^B.    \qquad
\end{eqnarray}

Here, $d\phi^I$ denotes the rotational Killing direction and $g_{AB}$ denotes the metric on the space $\mathcal{K}$. To remain in the most general setting as possible we allow a mixing (i.e a non-trivial dependence) of the $AdS_2$ metric and $\mathcal{K}$. The necessity for this will become obvious in the case of the Kerr-solution. \newline 
In the first step, we introduce a massive scalar field $\Phi$ in this background which obeys the Klein-Gordon equation. We decompose this into Kaluza-Klein towers by making the necessary conditions for the space $\mathcal{K}$. It is evident that by our symmetry assumptions it is reasonable to factor $\Phi = \Xi(R,T) \mathcal{Y}(y^A, \phi^I)$. \newline
Now, in the second step, we can write $\mathcal{Y} = e^{im^I \phi^I} Y(y^A)$ and $\Xi = \chi(R) e^{i\omega t}$. The Klein-Gordon equation takes the form $ \nabla_a \nabla^a \Phi - \mu ^2 \Phi = 0$. By the aforementioned decomposition, we can now proceed to separate the equations in question. We use furthermore, that the general covariant derivative in this space will lead us to the equation for the angular part after separation:

\begin{equation}
    \nabla_b \nabla^b (L(y)^2 \mathcal{Y}) - \mu^2 L^2(y) \mathcal{Y} - q^2 L^2(y) \mathcal{Y} = \lambda\mathcal{Y}.
\end{equation}

Here $\lambda$ is the separation constant between the radial and angular part. The constant $q$ which we will henceforth call charge $q$, can be understood in terms of the mixing between the space $\mathcal{K}$ and the space $AdS_2$ through a charged field which is implemented as a vector potential $A = -RdT$ and a charge $q = m_I k^I$. When we expand \ref{general}
we are left with the covariant derivative on $AdS_2$ as $\nabla_2$ and terms of the form $(\nabla_2 - iqA) \Phi$. \newline
The third step is now about finding the radial equation. This is straightforward due to the preparation we have made thus far. We arrive at 

\begin{equation}
    \label{radial}
    -\frac{1}{R^2} \frac{\partial^2 \Xi}{\partial T ^2}+ \frac{\partial^2 R \frac{\partial \Xi}{\partial R}}{\partial R} - \frac{2iq}{R} \frac{\partial \Xi}{\partial T} + (q^2  - \lambda - \mu^2) \Xi = 0.
\end{equation}

One example to show how general our approach is the Proca equation in this background to illustrate the universality of our approach. They stem from a Lagrangian density $\mathcal{L}$:

\begin{align}
    \mathcal{L} = \frac{F_{ab} F^{ab}}{4} + \frac{\mu^2}{2} A^a A_a
\end{align}

This gives rise to the Proca equation:

\begin{align}
    \nabla_a F^{ab} = \mu^2 A^a
\end{align}

where $A = dF$. Although gauge invariance is not given, one can quickly see that we have a Bianchi type $\nabla^a A_a = 0$. Now let $l^\mu$ be any null frame, then we make the ansatz \cite{reall} $A^a = \varphi_1 Z^a$ where $Z^a$ are vector harmonics on $\mathcal{K}$. The FKKS-Method \cite{Frolov:2018ezx} showed the separability of the Proca field on Kerr-AdS. If we write $A_i = F_{ab} l^a m^b_i$ where $l^a$ is on $\mathrm{AdS_2}$ and $m^b_i$ on $\mathcal{K}$, then we can get the equation:

\begin{equation}
    \left[\left(\nabla_2 - iqA\right)^2 + 1 - \lambda_1 \right] \varphi_1 = 0
\end{equation}

By finding $\lambda_1$ on $\mathcal{K}$, we can find the effective mass of the Proca field in this geometry. 

\subsubsection{Holographic Solution}

We now aim to solve this problem with an approach using the $AdS/CFT$ duality. We generically propose that that a one-dimensional conformal field theory is exactly dual to the scalar field. This dual field lives on the boundary as proposed by J.Maldacena. In order to do so, we aim to exploit the fact that generically the setup leads to a charged $AdS_2$ space. \newline
We wish to rewrite this in terms of more natural $AdS_2$ coordinates. For this, we introduce $\zeta = \frac{1}{R}$. This transforms (\ref{radial}) into (after inserting $\Xi = \chi(R) e^{i\omega t}$):

\begin{equation}
   \zeta^2 \frac{\partial^2 \chi}{\partial \zeta ^2} - L^2(y)(q^2 - \lambda - \mu^2) \chi + 2q\omega \zeta \chi + \zeta^2 \omega^2 \chi = 0
\end{equation}

This can be explicitly resolved in terms of Whittaker functions $W_{\kappa, \gamma}, M_{\kappa, \gamma}$. In order to understand how the dual CFT works, we need to compute the conformal scaling dimension at the boundary. For this, we need to find the asymptotic behaviour at the boundary. This is at $z \rightarrow 0$. There, we find for the Whittaker functions the result that, after asymptotically expanding:

\begin{equation}
    \chi \sim z^{-\Delta_{\pm}} 
\end{equation}

where $\Delta_{\pm} = \frac{1}{2} \pm \sqrt{q^2 - \mu^2 - \lambda + \frac{1}{4}}$ is the conformal scaling weight close to the boundary. \\

We introduce close to the horizon, a parameter $\eta = 1 - \frac{r_+}{r_-}$ which measures how close to extremality, i.e ($r_+ = r_-$) we are. We will later explain why two horiozons suffice, but in general this is because of the choice of de-Sitter space in our case. The connection to the extremal limit in the $\mathrm{AdS_2 /CFT_1}$ dual is made via the Hawking-temperature $T_H$ close to extremality, which is zero at extremality. Thus, via the surface gravity $\kappa$, we can write $T_H = \frac{1}{2\pi} \frac{d\kappa}{d\eta} \eta$. \newline
Finally, the quasinormal frequencies can be found canonically in a charged $AdS_2$ as the poles of the retarded Green's function in charged $\mathrm{AdS/CFT}$ (see \cite{retads2}): 

\begin{equation}
    \label{retrn}
    \mathcal{G}_R = (4\pi T)^{2 \Delta} \frac{\Gamma(2\Delta) \Gamma(\frac{1}{2} + \Delta - \frac{i\omega}{2\pi T} + iq_{\mathrm{AdS}}e_d) \Gamma(1 + \Delta - iqe_d)}{\Gamma(-2\Delta) \Gamma(\frac{1}{2} - \Delta - \frac{i\omega}{2\pi T} + iq_{\mathrm{AdS}}e_d) \Gamma(1 - \Delta - iqe_d)}. 
\end{equation}

The poles are located within ($e_d = L^2(y)$):

\begin{eqnarray}
        &\delta \omega = 2\pi T i(n - \frac{1}{2} + iq_{\mathrm{AdS_2}}L^2 - \Delta) \\ 
        &= 2\pi T \left( in - \frac{1}{2} + iq_{\mathrm{AdS_2}}L^2 - \left(\frac{1}{2} \pm \sqrt{q^2 - \mu^2 - \lambda + \frac{1}{4}} \right) \right).
\end{eqnarray}

At extremality, depending on the solution, we might have to also account for superradiance. This will be discussed in the Myers-Perry's solution.

\subsection{1-Cohomogeneity Myers-Perry's Solution}
\subsubsection{Solution in de-Sitter space}
We are in the following going to apply what we learned about the holographic approach to the Myers-Perry solution which exhibits an enhanced $U(n)$ symmetry (where $n$ is the number rotation axes) due to assuming equal angular momenta $a_i = a$. In Boyer-Lindquist coordinates ($r, t, \phi, \theta$) this looks like:

\begin{equation}
	\label{mpds}
	ds^2 = \frac{-f(r)}{h(r)} dt^2 + \frac{1}{f(r)} dr^2 + r^2 h(r) (d \psi + \mathcal{A} - \Omega(r) dt)^2  + r^2 d\Sigma^2. 
\end{equation}

Here the sphere $S^{N}$ is expressed as a fibration over $\mathbb{CP}^N $ and the helping functions $f(r) = 1- \frac{r^2}{L^2} - \frac{2M}{r^{2N}} \left( 1+ \frac{a^2}{L^2} \right) + \frac{2Ma^2}{r^{2N + 2}}$ and $h(r) = 1 + \frac{2Ma^2}{r^{2N+2}}$ \cite{Davey_2022}. This is a very common choice of coordinates for the Myers-Perry's solution \cite{cardoso2}. \newline
By Descartes rule of signs, the function $f(r)$ admits either 1 or 3 roots. Since a solution in de-Sitter space must admit at least 2 roots, $f(r)$ has 3 roots. The roots of the function $f(r)$ determine the locations of the horizons, which we name $r_+, r_-, r_c$. In the following, we are interested in the extremal case, namely when $r_+$ and $r_-$ coincide. This is equivalent to the case for which the Hawking Temperature of the black hole $T_H$ vanishes. \newline
The NH metric around the extremal case $r_+ = r_-$ can be re-expressed in terms of 

\begin{equation}
\begin{split}
    \label{NH}
    r \rightarrow r_+ + \epsilon R, t \rightarrow \frac{T}{\epsilon}, \\  \psi \rightarrow \Psi + \frac{\Omega(r_+)T}{\epsilon}
\end{split}
\end{equation}

where $\epsilon$ is an infinitesimal parameter. We can expand (\ref{mpds}) in terms of (\ref{NH}) and get the NH-expansion when dropping terms to first order in $\epsilon$ and dropping all terms proportional to $\epsilon$:

\begin{equation}
     ds^2 = -\frac{f''(r_+)}{2h(r_+)} R^2 dT^2 + \frac{2}{f''(r_+)} \frac{dR^2}{R^2} + r_+^2 h(r_+) \left(d\Psi + \mathcal{A} - R\Omega'(r_+) dT \right)^2 + r_+^2 d\Sigma^2 
\end{equation}

When naming $L^2 = \frac{2}{f''(r_+)}$ and rescaling $T,\Omega$:

\begin{equation}
        \Omega \rightarrow \Omega(r_+) \sqrt{h(r_+)} \\
        T \rightarrow L^2 \sqrt{h(r_+)} T,
\end{equation}

one can obtain the $AdS_2$ form of the (\ref{NH}) metric:

\begin{equation}
    \label{extrm}
     ds^2 = L^2 \left( -R^2 dT^2 + \frac{dR^2}{R^2}\right) + r_+^2 h(r_+)  \left(d\Psi + \mathcal{A} - R \Omega  L^2 dT \right)^2 + r_+^2 d\Sigma^2.
\end{equation}

When comparing the general approach to this specific example of a black-hole spacetime, we notice that we need to rewrite the $d\phi^I$ term as $d\Psi + \mathcal{A}$. Also, the fibration of $AdS_2$ with constant $L$ is truly a constant and independent of the coordinates on $\mathcal{K}$. The metric of this NH region can be written as proposed above as $\mathrm{AdS_2} \times \mathcal{K}$ where $\mathcal{K}$ is a squashed $(d-2)$-dimensional sphere with metric:

\begin{equation}
    ds^2 = (d\Psi + \mathcal{A})^2 + r_+ ^2 d\Sigma^2.
\end{equation}

With these preparations, we can find the radial equation of a scalar field $\Phi$ in this background. It is straightforward that the Klein-Gordon equation in charged $\mathrm{AdS_2}$ looks like:

\begin{equation}
    \label{mpdskge}
     R^2\chi''(R) + 2R\chi'(R) - \left[ \frac{(\omega - m\Omega L^2R)^2}{R^2} - L^2 \left( \mu^2 + \frac{\lambda}{r_+^2} + \frac{m^2}{r_+^2 h(r_+)}\right) \right] \chi(R) = 0. \qquad
\end{equation}

By comparing expressions, we note down that $q = m\Omega L^2$ and $A = -RdT$ as in the previous section. By comparing the differential equation to the previous section, just by inserting into $\Delta_{\pm} = \frac{1}{2} \pm \sqrt{q^2 - \tilde{\mu}^2 - \tilde{\lambda} + \frac{1}{4}}$ the frequencies, one can find $\delta \omega$. By using $\tilde{\lambda} =  \frac{\lambda}{r_+^2}, \tilde{\mu}^2 = \mu^2 + \frac{m^2}{r_+^2 h(r_+)}$, we arrive at:

\begin{eqnarray}
 \delta \omega^{NH}_{n,m} =  \left(i\left[ n + \frac{1}{2} + iq +  \frac{1}{2}\sqrt{4\Omega^2 L^4 m^2 - 4L^2 \left( \mu^2 + \frac{\lambda}{r_+^2} + \frac{m^2}{r_+^2 h(r_+)}\right) - 1} \right]\frac{d\kappa}{d\eta}\right) \eta \qquad \qquad
\end{eqnarray}

Since extremal rotating solutions are sourced by superradiance, $\omega^{(0)} < m\Omega_H$. Close to extremality, we can write the expression as:

\begin{eqnarray}
  \omega^{NH}_{n,m} =  m \Omega_H + \left( m \frac{d\Omega}{d\sigma} + i\left[ n + \frac{1}{2} + iq +   
 \frac{1}{2}\sqrt{4\Omega^2 L^4 m^2 - 4L^2 \left( \mu^2 +  \frac{\lambda}{r_+^2} + \frac{m^2}{r_+^2  h(r_+)}\right) - 1} \right] \frac{d\kappa}{d\eta}\right) \eta  \qquad \cr
\end{eqnarray}

This matches the result by \cite{Davey_2022}.

\subsection{Kerr-Solution}

The above discussion can be analogously applied to the Kerr solution of black holes. 
In this section,  we can finally explain the NH geometry of the 
Kerr solution.  In a paper by \cite{santoskn},  the NH region was closely examined in the classical context of GR. The NH geometry found can be locally described as a warped $AdS_3 \times \mathbb{CP}^n$.  Locally, the warped $AdS_3$ is that of $AdS_2$ fibred over $S^1$. This yields the local $SL(2,\mathbb{C}) \times U(1)$ symmetry.  This symmetry gives rise to left-moving $SL(2,\mathbb{C})$ and right-moving $U(1)$ Virasoro generators. \newline
In order to find the NH modes as before,  we can use the \textit{Poincare Patch} of the $AdS_2$ geometry. 
This can be expressed as:

\begin{equation}
 ds^2 = \left(\frac{1 + cos^2(\theta)}{2} \right) \left( -(1+y^2) d\tau^2 + \frac{dy^2}{1+y^2} + d\theta^2 \right) + \frac{2sin^2 (\theta)}{1 + cos^2(\theta)} (d\phi + y d\tau)^2
\end{equation}

If we use that a scalar massless mode can be separated as $\Phi = \psi(r,\theta) e^{im\phi + i\omega t}$,  then since the local isometry was $SL(2,\mathbb{C}) \times U(1)$ (and $SL(2, \mathbb{C})$ is the isometry group of $AdS_2$ space),  we can separate $\psi = \mathcal{Y}(y) \mathcal{T}(\theta)$.  This gives us then two separate ODEs for the $\theta$ and $y$ contribution respectively.  If we use slightly shifted coordinates that are no longer globally valid on $AdS_2$ but admit a coordinate singularity,  in this case \cite{kerstr}:
\begin{equation}
 ds^2 = \left(\frac{1 + cos^2(\theta)}{2} \right) \left( -(y^2) d\tau^2 + \frac{dy^2}{y^2} + d\theta^2 \right) + \frac{2sin^2 (\theta)}{1 + cos^2(\theta)} (d\phi + y d\tau)^2
\end{equation}

The massless KG equation for $\Phi$ is $\nabla_a \nabla^a \Phi = 0$.  After inserting the ansatz for $\psi$,  the ODE for $\mathcal{Y}$ reads:

\begin{equation}
y^2 \mathcal{Y}'' + 2y\mathcal{Y}' + \left( \frac{(\omega + my)^2}{y^2} + m^2\right) \mathcal{Y} = 0
\end{equation}

The Whittaker function as earlier allows us to describe its asymptotics by polynomials close to the horizon.  This allows to read off the conformal weight as:

\begin{equation}
	\Delta_\pm = -\frac{1}{2} \pm \sqrt{K - 2m^2 + \frac{1}{4}}
\end{equation}

By reinserting into \ref{retrn} and using the $\Gamma(-n) = 0$ prescription,  we can identify the NH modes as for the MPdS case as:

\begin{equation}
	\omega = m\Omega_H + \left( m\frac{d\Omega}{d\eta} + i\left( n + \frac{1}{2} + im + \sqrt{K - 2m^2 + \frac{1}{4}} \right) \right) \eta
\end{equation}

This matches our result and especially now only requires a discussion of the parameter $K$. This is a solution of the differential equation \cite{santoskn}:

\begin{equation}
    \label{distort}
    \mathcal{T}''  +\cot(\theta) \mathcal{T}' + \left(K - \frac{m^2}{\sin^2 (\theta)} + \frac{m^2}{4} \sin^2(\theta) \right)\mathcal{T} = 0.
\end{equation}

If $m = 0$, $\mathcal{T}$ will is an azimutally symmetric spherical harmonic $Y_{l0}$ and, thus, $K = l(l+1)$. If, however, $m \neq 0$, then the solutions to \ref{distort} need to be found numerically. $K$ is then the eigenvalue of the corresponding differential operator. \newline

We hence showed that we can systematically find the charge of the scalar field in the NH region and then employ the retarded Green's function method by \cite{son} in order to determine the QNM frequencies. In the extremal limit, the AdS/CFT conjecture hence remains an important tool that gives many results which align with the classical predictions.

\subsection{The CFT Dual}

Since we are using the bulk/boundary duality, the aforementioned calculation provides us with insights about the dual field theory and its constituents. The metric is that of a warped $AdS_3$ space, which we want to find dual to a two-dimensional $\mathrm{CFT_2}$. We work with the local symmetries of the NH-region that are as $SL(2,\mathbb{C})_L \times \left( U(1) \times U(1)\right)_R$ which then give rise to leftmoving and rightmoving generators. The resulting leftmoving Virasoro generators $L_n$ (for $n \in \mathbb{Z}$) can be used to span the set of conformal generators in this case. We know that the dual is governed by the dilaton operator $D = L_0 + \bar{L}_0$ (left- and rightmoving) with eigenvalue $\Delta = h_L + h_R$ of the dilaton operator. Hence, we expect the CFT dual to the bulk field to propagate with propagator:

\begin{align}
    G(x,y) \sim \frac{1}{|x-y|^{2\Delta}}.
\end{align}

The main parameters that will determine the properties of the CFT itself are determined by the Bekenstein-Hawking entropy. In the case of the Myers-Perry's solution, this means: 

\begin{align}
    S_{H} = \frac{k_B A}{4l_P^2} \sim \frac{\pi}{2}\int_{0}^{\pi} d\theta \sqrt{r_+^4 h(r_+) L^2}.
\end{align}

We can use the \cite{frolov} approach to find the left-moving temperature and construct the Virasoro generators via the Hawking Bekenstein entropy. We note (which is a well-known fact) that the entropy of the black-hole is not zero at extremality whereas the Hawking temperature is. 
We first work in the near-horizon regime of the MPdS solution. We recall that we derived $T_H = \frac{1}{2\pi} \frac{d\kappa}{d\eta} \nu$ where $\eta = 1-\frac{r_+}{r_-}$ is a parameter leading to the extremality limit. If we introduce $T_H' = \frac{\partial T_H}{\partial r_+} \big|_{r_+ = r_-}$, we can define with $\Omega_H' = \frac{\partial \Omega_H}{\partial r_+} \big|_{r_+ = r_-} $, the left-moving temperature as $T_L = \frac{T'_H}{\Omega'_H}$. We can use Cardy's formula to postulate that $S = \frac{\pi^2}{3} c_L T_L$. In order to find the central charge, we now need $T_L$. This can be found in two ways. The more geometric approach is to consider what happens to the dual CFT when going from the zero-temperature case to the finite temperature one is to map it to a torus with radii $\frac{1}{T_{L/R}}$. From this, we can derive that $ T_L = \frac{1}{2\pi \Omega L^2}$ and $c$ can then be read off. The other approach would be to just explicitly calculate the derivatives of $T_H, \Omega_H$ and take the respective extremal limit. Thus in the (NH) metric, we can find:

\begin{equation}
    c_L = \frac{3S}{T_L} = 6r_+^2 L^3 \Omega \pi^2.
\end{equation}

In general the Frolov-Thorne temperature \cite{Perry_2022}, can be found as in the explanation before and leads in $N$ dimension to the result and find. In the case of the Myers-Perry solution (not only in NH space), we can derive the extremal central charge as:

\begin{equation}
    T_L = \frac{1}{2\pi}\frac{f'' (r_+)}{2r_+ h(r_+)} \frac{(r^{2N+1})^2}{(2N+2)\frac{h(r_+)}{r_+} + h'(r_+)} = \frac{f''(r_+) (r_+^{2N+2})}{(2N+1)2h(r_+)}.
\end{equation}

Here, one needs to use that the surface gravity vanishes at extremality. This is given by $\kappa = \frac{|f'(r_+)|}{2r_+ h(r_+)} = 0$. This fixes the central charge again via Cardy's formula. 
\newline
The local $SL(2,\mathbb{R})$ geometry gives rise to Virasoro generators that generate the conformal symmetries of the space. The construction in this case is going to be similar to the Kerr-Case considered in \cite{Hartman_2009}. The construction follows a well-known construction in literature of the \textit{Asymptotic Symmetry Group}. The local generators are constructed as in the Kerr/CFT \cite{strominger} and will be omitted here. We note that the left moving generators $L_n$ satisfy a Virasoro algebra with central extension $c_L$ whereas the right-moving $\mathfrak{l}_n$ do not have a central extension term, because the charge $c_R = 0$ at extremality. The calculation is straightforward because $T_R = 0$. Close to extremality, $c_R \neq 0$ because the temperature $T_R$ in only $0$ in the strict extremality limit. \newline
One of the interesting insights provided by the use of the AdS/CFT duality is that of using knowledge about the dual CFT to say something about the stability of the scalar field in this NH region. For this the Breitenloher-Friedman bound is useful, as it tells us necessary criteria for the bulk field to be renormalisable. \newline
The Breitenloher-Friedmann bound states $m^2 L^2 > -\frac{d^2}{4}$. For 1-cohomogeneity solutions of the Myers-Perry's black hole, we know:

\begin{equation*} 
\Delta_{\pm} = \frac{1}{2} \pm \sqrt{q^2 - \mu^2 - \lambda + \frac{1}{4}} = \frac{1}{2} \pm \frac{1}{2}\sqrt{4\Omega^2 L^4 m^2 - 4L^2 \left( \mu^2 + \frac{\lambda}{r_+^2} + \frac{m^2}{r_+^2 h(r_+)}\right) - 1}.
\end{equation*}

Specifically, this means that we can find a stability bound on scalar perturbations in an $AdS_2$ spacetime. Since the effective mass is $\sqrt{q^2 - \mu^2 - \lambda + \frac{1}{4}}$, we get an imaginary mass if:

\begin{equation}
    q^2 +\frac{1}{4} \leq \mu^2 + \lambda.
\end{equation}

This is the $\mathrm{AdS_2}$-Breitenlohner Freedman bound in $\mathrm{AdS_2}$ space. 

\section{Fermionic Modes}

\subsection{General Dirac Fields}
\subsubsection{Setup}
\qquad
In this section, we aim to explain how to understand fermionic perturbations in the NH region of extremal black holes. For this, we need to first describe the necessary semantic tools which we need in order to make our theory rigorous. We work in the same background as before. 

\begin{eqnarray*}
\label{general}
 ds^2 = L(y)^2 \left( -R^2 dT^2 + \frac{1}{R^2} dR^2 \right) +  
        g_{IJ} (y) (d \phi^I - k^I R dT)  
         (d \phi^J - k^J R dT) +  g_{AB}(y) dy^A dy^B.    \qquad
\end{eqnarray*}

Spinor fields live in representations of the Lorentz-group $SO(3,1)$ of half-integer spin. Under a Lorentz transformation $x_{\mu} = \Lambda^{\mu}_{\nu} x^\nu$, the transformations of the Lorentz group act in blocks of generators $\mathcal{M}_{\alpha \beta}$, such that a finite Lorentz transformation looks like $\Lambda = \exp(i\theta_{\alpha \beta} \mathcal{M}^{\alpha \beta})$. The parameters $\beta_{\alpha \beta}$ act as boost parameters and are totally antisymmetric by construction of the generators of the Lorentz group. If $\Psi$ denotes a spinor field, and $S[\Lambda]$ is the spinor representation of $SO(3,1)$, under a finite Lorentz transformation the spinor transforms as $\Psi \rightarrow S[\Lambda] \Psi(\Lambda^{-1} x)$.  \newline
If we wish to consider fields with higher spin than scalar fields, it is central to introduce a vielbein basis of the background one works with. We refer to a frame of covectors $e^{a} = e^{a} _{\mu} dx^\mu$. These satisfies the orthonormality relation $g^{\mu \nu} e^a_{\mu} e^b _\nu = \eta^{ab}$, where $\eta^{ab}$ is the Minkowski metric. Thus the vielbein frame endows the spacetime manifold $\mathcal{M}$ with its Lorentzian structure. The dual basis lowers the latin indices $a,b$. In order to be able to describe spinor fields, we need to introduce the gamma matrices $\gamma^a$ which we will use in the Weyl representation. We will be working here in a setup with at most 5 dimensions, and so the spinor dimension will be 4. \newline
The last formal ingredient is the so called spin connection. This enables us to consider quantum fields with higher spin $s$ (here for $s = 1/2$). For this we define  a general covariant derivative:

\begin{equation}
    \mathcal{D_\mu} = \partial_\mu - \Gamma_\mu - iqA_\mu 
\end{equation}

where $\Gamma_\mu$ is the spin connection. We define the spin connection via \\ 
$\gamma_{cab} = e^\mu_c \left( \partial_\nu e_{a \mu} -\Gamma_{\mu \nu}^{\beta} e_{a \beta}\right)e_{\beta}^\nu$. Intuitively, the spin connection is the connection such that the frame $e^a _\mu$ is covariantly conserved: $\nabla_a e^{c}_\mu = 0$. Thus the formal abstract definition: $\omega_{\mu ab} = e^{\nu}_a \nabla_{\mu} e_{b\nu}$ can be made as well. With this the expression for $\Gamma_{\mu}$ can be written as:

\begin{equation}
    \Gamma_\mu = -\frac{1}{2} e^{c}_{\mu} \gamma_{cab} S^{ab} = -\frac{1}{2} e^{c}_{\mu} \lambda_{cab} \frac{1}{2} [\gamma^a, \gamma^b].
\end{equation}

Here $S^{ab}$ is the spinor generator of Lorentz transformations. As a last remark one can point out that the $e^a_\mu$ act as projectors between the Minkowski latin indices and the greek indices in curved spacetime. Hence $\gamma^\mu = e^{\mu}_a \gamma^a$ and $\partial_a = e_a^{\mu} \partial_\mu$. 

\subsubsection{Dirac Equation in Curved Space}

The action for a fermion coupled to a curved background can be written as:

\begin{equation}
    \label{dirac}
    S_D = \int d^4 x \sqrt{-g} \left( [\bar{\Psi} \gamma^{\mu} (\mathcal{D}_\mu \Psi) - (\mathcal{D}_\mu \bar{\Psi}) \gamma^{\mu}  \Psi ] - \mu \bar{\Psi} \Psi \right)
\end{equation}

where $\bar{\Psi}$ is the Dirac adjoint. By varying the action w.r.t. $\bar{\Psi}$, we obtain the Dirac equation i$\gamma^{\mu} \mathcal{D}_{\mu} - \mu\Psi = 0$. Since we chose the Weyl representation of the four-spinor, the spinor field naturally decomposes to two chiral left- and right-handed bispinors. We can use the projectors $P_{\pm} = \frac{1 + \gamma^5}{2}$ to project to either side.\newline
We write:

\begin{align}
    \gamma_0 = 	\begin{bmatrix} 
	0 & iI_2 \\
	iI_2 & 0\\
	\end{bmatrix}, 
 \gamma_1 = \begin{bmatrix} 
	0 & i\sigma_3 \\
	-i\sigma_3 & 0\\
	\end{bmatrix}, 
  \gamma_2 = \begin{bmatrix} 
	0 & i\sigma_1 \\
	-i\sigma_1 & 0\\
	\end{bmatrix}, 
  \gamma_3 = \begin{bmatrix} 
	0 & i\sigma_2 \\
	-i\sigma_2 & 0\\
	\end{bmatrix}
\end{align}

We now aim to write (\ref{dirac}) explicitly in this general background and then solve for the frequencies in the same way we solved for the scalar mode frequencies. We can construct a co-frame basis:

\begin{eqnarray*}
    &e^{0} = -LR  dT  \qquad
    &e^{1} = \frac{L}{R}dR \\
    &e^2 =  e_{I} (d\phi^I - k^I R dT) \qquad
    &e^i = r_+ \hat{e}^i_adx^a
\end{eqnarray*}

For the Myers-Perry's solution we will use $d\phi^I = d\Psi + \mathcal{A}$. The $\hat{e}^i_a$ represents the vielbein on the Fubini study metric $\hat{g}_{ab}$ in  the NH limit. When we lower indices, we then get:

\begin{equation}
    e_0 = \frac{1}{LR} \left( \partial_T - \Omega L^2 R \partial_{\Psi} \right) \qquad
    e_1 = \frac{R}{L} \partial_R \qquad e_2 = \frac{1}{r_+ \sqrt{h(r_+)}} \partial_\Psi
\end{equation}

and on $\hat{e}^i_a$, we need to use an orthogonalisation procedure:

\begin{equation}
    e_i = \frac{1}{r_+} \left(\hat{e}_i - \langle \mathcal{A}, \hat{e}_i \partial_\Psi \rangle \right)
\end{equation}

where $\hat{e}_i$ is the dual basis to $\hat{e}^i$. The spin connection $\omega$ stems from the principle bundle structure associated with the spin group $\mathrm{Spin(n)}$. Thus, we can compute its coefficients using the Cartan Structure Equation $de + \omega \wedge e = 0$. 

\begin{equation}
     \gamma^\mu \Gamma_\mu = e^{\mu}_a \gamma^a \Gamma_{\mu} = -\frac{1}{2}e^{\mu}_a \gamma^a \omega_{\mu (ab)}[\gamma^a, \gamma^b] =   -\frac{1}{2}e^{\mu}_d e^c_\mu \gamma^d \lambda_{abc} [\gamma^a, \gamma^b].
\end{equation}

As well as $\gamma^\mu \partial_\mu $:

\begin{equation}
    \gamma^\mu \partial_\mu  = e^{\mu}_a \gamma^a \partial_\mu = \gamma^a \partial_a. 
\end{equation}

Thus, the Dirac equation looks like: 

\begin{equation}
    \left(\gamma^a \partial_a \ -\frac{1}{2}e^{\mu}_d e^c_\mu \gamma^d \lambda_{abc} [\gamma^a, \gamma^b] \right) \Psi = m \Psi.
\end{equation}

The spin connection coefficients are then:

\begin{eqnarray*}
        \omega_{01} = -\frac{e^0}{L} - \frac{r_+ \sqrt{h(r_+)} e_2}{2} \qquad 
        \omega_{02} = \frac{r_+ \sqrt{h(r_+)} e_1}{2} \qquad &\omega_{0i} = 0 
\end{eqnarray*}

\begin{equation}
    \omega_{12} = \frac{r_+ \sqrt{h(r_+)} e_0}{2} \qquad \omega_{1i} = \frac{R}{Lr_+} e^i \qquad \omega_{2i} = \frac{\sqrt{h(r_+)}}{r_+} \mathcal{J}_{ij} e^j
\end{equation}

\begin{equation*}    
    \omega_{ij} = -\frac{\sqrt{h(r_+)}}{r_+} \mathcal{J}_{ij} e^2 + \frac{1}{r_+} \hat{\omega}_{kij} e^k
\end{equation*}

Here $\mathcal{J} = -\frac{1}{2}d\mathcal{A}$ is the Kähler-form on $\mathbb{CP}^n$. We now write down $\Gamma_a$:

\begin{equation}
    \gamma^0\Gamma_0 = \frac{1}{2} \gamma^1 \left( \frac{1}{L} + \gamma^0\gamma^2\frac{r_+ \sqrt{h(r_+)}}{2} \right) \qquad \gamma^1 \Gamma_1 = -\frac{1}{2} \hat{\gamma}^2\left( \frac{r_+ \sqrt{h(r_+)}}{2}\right)
\end{equation}
\begin{equation*}
    \gamma^2 \Gamma_2 = -\frac{1}{4} \hat{\gamma}^2 \left( -\frac{r_+ \sqrt{h(r_+)}}{2} - \frac{\sqrt{h(r_+)}}{r_+} \mathcal{J}_{ij}\right)
\end{equation*}

We introduced the notation that $\hat{\gamma}^2 = \gamma^0 \gamma^1 \gamma^2$. By expressing the equation in terms of bispinors $\Psi_L, \Psi_R$:

\begin{equation}
\label{sepdir}
        \Psi_L = \mathcal{N} \exp{(i\omega T +im_I\phi^I)}\begin{bmatrix}
           R_1(R) Y_1(y)\\
           R_2(R) Y_2(y)
         \end{bmatrix}
\end{equation}

the Dirac equation takes the form (for one component):

\begin{equation}
    \frac{-i}{RL} \left( i\omega - i\Omega L^2 Rm \right) (R_1 Y_1) - i\frac{R}{L} \left( \partial_R  \right)(R_1 Y_1) + i\frac{1}{r_+ \sqrt{h(r_+)}} m (R_2 Y_2) + \frac{1}{r_+}\hat{\mathcal{D}}(R_2 Y_2) + \frac{1}{2L} R_1 Y_1= 0\end{equation}

We are interested in spinor harmonics on $\mathbb{CP}^n$, hence: $\hat{\mathcal{D}}Y_{1/2} = \lambda Y_{1/2}$. This gives us the radial equation for $R_1$ (or $R_2$ respectively after separating), namely that it follows:

\begin{equation}
    R \partial_R ( \left( R \partial_R \right) R_1) - \left[\frac{(\omega - \Omega L^2 m R)^2}{R^2} - \frac{i\omega}{R} + \frac{1}{4}- L^2 \left( \mu^2 + \frac{\lambda^2}{r_+^2} + \frac{m^2}{r_+^2 h(r_+)} \right)\right] = 0
\end{equation}

This can be again written very abstractly as the charged scalar field equation on $\mathrm{AdS_2}$ where we have called the scalar field $\chi_{1/2}$ to show the explicit spin dependence. Here $A = - RdT$ is again the gauge field in this background with charge $m \Omega L^2$. We can solve this with similar methods as in Section 2. 
 
\begin{equation}
   \left((\nabla_2 - iqA)^2  -  \mu^2 +\frac{1}{4} \right)\chi_{1/2} = \lambda \chi_{1/2}
\end{equation}

\subsection{Holographic Methods}

Before we can analyse the quasinormal behaviour on $AdS_2$, one needs to impose the correct boundary conditions that will make the $AdS/CFT$ duality physically reasonable. Since we are describing quantum fields on the asymptotic boundary, and this boundary is described purely by its radial characteristics, we are interested in the pure derivative terms on the boundary that act as our boundary conditions. Precisely if we consider $\delta S$ only, we neglect the boundary action $\delta S_{bdry}$. We vary our action such that $\delta S = 0$, but not $\delta \left( S + S_{bdry} \right) = 0$. \newline
In the context of the AdS/CFT duality we claim that a \textit{well-defined} conformally symmetric quantum field which dual to the Dirac spinor $\Psi$ resides on the boundary. Since we are interested in the quasinormal modes of such fermionic perturbations, we are interested in \textit{ingoing} asymptotic boundary conditions. In the following, we are going to work within these assumptions to understand the field theoretic approach to this. \newline
Specifically, we recall we found a $\Delta_{\pm}$ which corresponds to two different conformal weights for scalar fields $\phi = \phi_+ + \phi_-$. Similarly, we wish to decompose $\Psi = \Psi_+ + \Psi_-$. Since the behaviour at the boundary is asymptomatically determined by its scaling for $r \rightarrow \infty$, we are interested in the behaviour of the radial $\gamma^1$ far away. Indeed it is advantageuos to propose the ansatz for the boundary conditions:

\begin{equation}
\label{sepdir2}
        \Psi_\pm = \mathcal{N} \exp{(i\omega T + im_I\phi^I)}\begin{bmatrix}
          i (\chi_1(R) \pm i\chi_2(R)) S_1(y)\\
           ( \chi_1(R) \pm i\chi_2(R) )S_2(y) \\
            ( \chi_1(R) \pm i\chi_2(R) )S_1(y) \\
           -i( \chi_1(R) \pm i\chi_2(R) )S_2(y)
         \end{bmatrix}
\end{equation}

This allows to differ between the different conformal dimensions associated to the dual fields $\Delta_{pm}$.
By separating and obtaining the conformal scaling dimension of the field, we can then use the charged Green's function approach to explicitly find the quasinormal frequencies. \newline

\subsubsection{Finding Asymptotic Scaling}

We conclude this section by briefly describing the parameter $\lambda$ which we need for the conformal dimension. Then, we can explain the value of $\Delta_{\Psi}$. The eigenvalue can be explicitly found as the spinor harmonic on $\mathbb{CP}^n$. The strategy to find $\lambda$ follows the idea that $\mathbb{CP}^n$ is a homogeneous space $SU(n+1)/(U(n))$. The explicit computation now uses Lie theory \cite{spinor} to find the eigenvalues:

\begin{align}
    \lambda^2 = l^2 + \frac{1}{2} l\left(3n -2k - 1\right) + \frac{1}{2} (n-k) (n-1)
\end{align}
\begin{align}
    \beta^2 = l^2 + \frac{1}{2} l\left(3n -2k + 1\right)  + \frac{1}{2} (n-k) (n+1)
\end{align}

A consistency check reveals for $N = 1$, that $\lambda^2 = \beta^2 = l(l+1)$. Thus, we can state the spinor conformal dimension $\hat{\Delta}_\Psi$:

\begin{align}
    \hat{\Delta}_\Psi = \frac{1}{2} \pm \sqrt{\frac{1}{2} - L^2\left(\frac{\lambda}{r_+^2} + \frac{m^2}{r_+^2 h(r_+)} + \mu^2 \right) + m^2\Omega^2L^4}
\end{align}

Now, we can similarly find the QNF of the Dirac Modes as:

\begin{eqnarray}
\label{diracq}
  \omega^{NH}_{n,m} =  \left(  i\left[ n + \frac{1}{2} + iq +   
 \sqrt{\frac{1}{2} - L^2\left(\frac{\lambda}{r_+^2} + \frac{m^2}{r_+^2 h(r_+)} + \mu^2 \right) + m^2\Omega^2L^4} \right] \frac{d\kappa}{d\eta}\right) \eta  \qquad \cr
\end{eqnarray}

Unlike scalars fermions do not obey a superradiant bound, which explains the missing of these terms in \ref{diracq}.

\subsection{Fermionic CFT Dual}

We can use the aforementioned results to discuss the CFT dual to a fermionic field theory close to the horizon. After having discussed the CFT dual to a bosonic scalar field, we will need to understand the behaviour of fermions in local $AdS_2$. This is particularly relevant if one checks the behaviour of fermions in global $AdS_2$ for any possible supersymmetric extensions of the boundary CFT. In general the results about the conformal dimension:

\begin{equation}
    \Delta_{\Psi} = \frac{d}{2}  + |m_{\mathrm{eff}}| = \frac{1}{2} \pm \sqrt{\frac{1}{2} + \Lambda}.
\end{equation}

This obeys the $AdS_2$ Breitenlohner-Freedman bound to give:

\begin{equation}
    \Lambda \geq -\frac{1}{2}.
\end{equation}

Since certain combinations of the quantum parameters render the conformal scaling dimension complex, we find that the effective mass in AdS is subject to \textit{tachyonic} instabilities. This sets mass bounds on the fermions in the bulk and thus suggests that fermions behave differently from bulk bosons. \newline

The propagator of fermions in charged $AdS_2$ space can be found as \cite{retads2}:

\begin{equation}
        \mathcal{G}_R = (4\pi T)^{2 \Delta} \frac{\Gamma(2\Delta) \Gamma(\frac{1}{2} + \Delta - \frac{i\omega}{2\pi T} + iq_{\mathrm{AdS}}e_d) \Gamma(1 + \Delta - iqe_d)}{\Gamma(-2\Delta) \Gamma(\frac{1}{2} - \Delta - \frac{i\omega}{2\pi T} + iq_{\mathrm{AdS}}e_d) \Gamma(1 - \Delta - iqe_d)} \times \frac{m - iqe_d - \Delta}{m- iqe_d + \Delta}
\end{equation}

When considering the near-extremal region, we need to consider a finite dimensional generalisation of the CFT. We make a change of coordinates to $t = \frac{1}{2\pi T} \exp(2\pi T \tau)$:

\begin{equation}
    \langle \mathcal{O}(\tau) \mathcal{O}(\tau') \rangle = \left(\frac{d\tau}{dt} \right)^{1+\Delta} \left(\frac{d\tau'}{dt'} \right)^{1+\Delta} \langle \mathcal{O}(t) \mathcal{O}(t') \rangle
\end{equation}

As \cite{retads2} writes about charged AdS space, the fermionic perturbations close to the horizon of an extremal black hole behave as a Fermi fluid. Exploiting this in the case of the rotating black hole, in the finite temperature generalisation, the gauge field $A = RdT = \frac{1}{\zeta} dT = \frac{1}{\zeta} d\tau + d \tanh(\zeta/\zeta_0)$ where $\zeta_0$ is such that $T = \frac{1}{2\pi \zeta_0}$. Then, we can write for the two-point correlation function:

\begin{equation}
    \langle \mathcal{O}(\tau) \mathcal{O}(\tau') \rangle = \left(\frac{\pi T}{\sinh(\pi T(\tau - \tau'))} \right)^{1 + 2\Delta} \exp(-2\pi iT(m\Omega L^2) (\tau - \tau')).
\end{equation}

In the case of an oscillatory solution close to the horizon, we observe that om the extremality region, the dicrete quasinormal modes of the retarded Green's function form a branch cut. If we are close to extremality, the poles begin accumulating at $\omega = 0$. This leads to (in the \textit{oscillatory case} only) to a \textit{Fermi Sea} close to the horizon.

\newpage
\section{Discussion and Conclusion}
In this article, we examined the Quasinormal Frequencies (QNFs) using methods from the holographic duality principles and used them to understand bosonic as well as fermionic modes in the near horizon region of extremal rotating black holes. The renormalisability criteria provided us with stability criteria on the effective masses of the dual boundary operators that allowed us to describe these perturbations also in light of the dictionary between CFTs and their bulk duals. We were able to compute the central charges of the theory for the case of the Myers-Perry's black hole with equal angular momenta and the Hawking temperature as well as the Hawking entropy. With this, it was then possible to describe the boundary CFT as an ultimately thermal theory in this limit. 

In the case of spin-2 fields, we wish to analyse the behaviour due to the holographic dictionary that we have proposed thus far. If we start from the Fierz-Pauli action, we can find the perturbation operator for the Myers-Perry black hole via varying the action and rewriting the equation of motion in this case. If we proceed to decompose the spin 2 field $h_{\mu \nu} = \chi (r) \mathcal{Y}_{\mu \nu}$,  we can find the angular perturbation operator very similarly. Its eigenvalues are $\lambda_{nlm}$. We can extract the frequencies as per earlier to get $\omega_{nlm}$.  \newline
The claim we make now is that the quantum dual system behaves very similarly to the system which we found before. In general each of the operators dual to one such eigenmode $\omega_{nlm}$ will be dual to the correlation function of the two quantum operators on the boundary. When comparing the frequencies we found, we see that the spectrum is very similar in structure to the fermionic Matsubara frequencies and so the Fermi-Dirac distribution supported by a chemical potential $\mu$ would be a good fit for the parameters. \newline
One phenomenon that we wish to consider is the so-called thermal ringdown process of the boundary CFT. In this case, for example in black hole mergers, we would observe the superposition of many such normal modes with eigenfrequencies $\omega_{nlm}$. If we can write the mode $\tilde{h}_{lm} (\omega)$ in Fourier space as:

\begin{equation}
    \tilde{h}(\omega) = \frac{1}{2\pi} \int dt h_{QNM, lm} (t) e^{i\omega t} = \frac{i}{2\pi} \sum_{n} \frac{C_{lm}}{\omega - \omega_{lmn}} \exp({i\omega t^*}).
\end{equation}

If we now claim, indeed, that the statistical behaviour of $|\tilde{h}(\omega)|^2$ is governed by a Fermi Dirac distribution:

\begin{equation}
    \frac{1}{\exp\left(\frac{(\omega - m\Omega_H)}{T}\right) + 1}
\end{equation}

, for near extremal black holes with $T_H \sim \frac{d\kappa}{d\sigma} \sigma$, then we can understand something about the thermal ringdown of the boundary CFT. If during a merger process the black-hole rings, i.e we are led to excited overtones, this is akin to excited states of the dual quantum operators. Recent analysis of GW has shown this to be the case \cite{narita}. What this additionally tells is us is the behaviour of the long time correlation function of the quantum dual which we can understand as:

\begin{equation}
    \langle \mathcal{O}_{l,m} \mathcal{O}_{l,-m} \rangle = \sum_n \exp({-i\omega_{lmn}} \Delta t).
 \end{equation}

 Here we take the ensemble average in $\langle \mathcal{O}_{l,m} \mathcal{O}_{l,-m}\rangle$. Hence, this provides new insights into the holographic thermalisation process as a whole.\newline
 
 In the case of the Myers-Perry's solution this is akin to using the angular eigenvalue for a spin-2 field. This can be derived from separation of the equation of motion:

 \begin{equation}
    \Box h_{ab} + 2R_{acbd} h^{cd} = 0.  
 \end{equation}

The eigenvalue is (\cite{Durkee_2011}):

\begin{equation}
    \frac{\lambda_{lm}}{L^2} = 4(1-\sigma) \left( \frac{N}{r_+^2} + \frac{N+1}{l^2} \right) +\frac{4\kappa (\kappa + N)}{r_+^2}
\end{equation}

Hence, we can write the frequencies $\omega_{lmn}$:

\begin{equation}
    \omega_{lmn} = m\Omega_H + m\frac{d\Omega}{d\sigma} \sigma +  i\left(\frac{1}{2} + n \pm \delta_{lm} \right)\frac{d\kappa}{d\sigma}\sigma
\end{equation}

where $\delta_{lm}$ is part of conformal dimension of this tensor $\delta_{lm} = \sqrt{\frac{1}{4} + \lambda_{lm}}$ which we just abbreviate. We quickly see since $\lambda_{lm}$ is non-negative, the AdS-BF is definitely satisfied and allows a well-defined dual boundary operator. So if we wish to use the above stipulation it can definitely be valid to write:

\begin{equation}
    \label{overtone}
    \langle \mathcal{O}_{l,m} \mathcal{O}_{l,-m} \rangle = \sum_n \exp({-\delta \omega_{lmn}} \Delta t) = \sum_n \exp \left[-i\left(i\frac{1}{2} + n  \pm \delta_{lm} \right)\frac{d\kappa}{d\sigma}\sigma \Delta t \right]
 \end{equation}

 Hence, the overtone modes of rotating black holes close to extremality can provide with valuable insights about the boundary correlators. Explicitly, what, would be interesting is to understand the predicted mass spectrum of (\ref{overtone}). This would allow us to examine the holographic duality principles in a more realistic environment. \newline
 Another aspect that is interesting is the aspect of Strong Cosmic Censorship in the formulation of Christodulo. For uncharged rotating black holes it has been shown by \cite{dafermos} that this hypothesis fails to hold. For near-extremal charged black holes there is a strong hint it might be violated.  This gives us the possibility to calculate the spectral gap $\alpha$ to investigate the strongest version of it by Christodulo. Here $\alpha$ is the minimum of $\Im(\omega)$.  Close to extremality $\kappa_- = -\frac{d\kappa}{d\eta}$ and we define $\beta = -\frac{\alpha}{\kappa_-}$. For the MP-dS black hole, this means for $\beta$: 

\begin{equation}
    \label{beta}
     \beta = \left[\frac{1}{2} + \frac{1}{2}\sqrt{4\Omega^2 L^4 m^2 - 4L^2 \left( \mu^2 + \frac{\lambda}{r_+^2} + \frac{m^2}{r_+^2 h(r_+)}\right) - 1} \right]
\end{equation}

If $\beta \geq \frac{1}{2}$, then the field is in $H^1_{loc}$ and then the Christodulo version doesn't hold for the near-horizon modes close to extremality. This region is however, crucial because we are interested in the value of $\beta$ in this region specifically. We see here in (\ref{beta}) that then \textit{a priori} $\beta \geq \frac{1}{2}$.  We find that the Strong Cosmic Censorship hypothesis not violated if the $\mathrm{AdS_2}$-BF bound is respected. Thus, if the bound on $\beta$ should be respected, then we are requiring a violation of the AdS-BF bound. In (\ref{beta}), we need choose $m$ large enough such that the bound is violated. By finding only one family of such modes that satisfy the $\beta$ bound, we satisfy the SCC criterion. This can then be explored numerically. \newline
We can similarly discuss this for the case of Dirac fermions. For scalar modes it was proven that this holds by \cite{Christodoulou:2008nj}. We believe it to be plausible that since, we decomposed the spinor into its scalar ingredients, for this to hold as well in this case. This leaves us with the same condition, namely $\beta < \frac{1}{2}$ for the SCC to hold. For the Reissner-Nordström case it was proven in \cite{Destounis_2019} that the perturbations lie in $H_{loc}^s$ for $s < \frac{1}{2} + \frac{\alpha}{\kappa_-}$. Hence, the definition of $\beta$ can be justified very analogously. We find for $\beta_{Dirac}$:

\begin{equation}
    \beta_{Dirac} = \frac{1}{2} + \sqrt{\frac{1}{2} - L^2\left(\frac{\lambda}{r_+^2} + \frac{m^2}{r_+^2 h(r_+)} + \mu^2 \right) + m^2\Omega^2L^4}.
\end{equation}

Again we would need as a \textit{necessary} condition that the $\mathrm{AdS_2-BF}$ to be violated. This again makes a striking connection for near-extremal black holes and dual conformal structure. 
\newpage

\bibliographystyle{unsrtnat}
\bibliography{jhepexample}
\end{document}